\documentclass{ifacconf}
\usepackage{amsmath,amssymb,amsfonts}
\newcommand\numberthis{\addtocounter{equation}{1}\tag{\theequation}}
\makeatletter
\usepackage{graphicx}      
\graphicspath{ {./Figures/} }
\usepackage{comment,url}
\usepackage{caption}
\usepackage{subcaption}
\usepackage{xcolor,bm}
\usepackage{diagbox}
\usepackage{algorithmic}
\usepackage{textcomp}

\usepackage{natbib}        
\begin{document}
\allowdisplaybreaks

\begin{frontmatter}

\title{Bearing-Only Solution to the Fermat-Weber Location Problem for Unicycle Agent}

\author[First]{Hong Liang Cheah} 
\author[First]{Mohammad Deghat} 
\author[First]{Jose Guivant}

\address[First]{School of Mechanical and Manufacturing Engineering, University of New South Wales (UNSW), 2052, NSW, Australia  (e-mail: h.cheah@unsw.edu.au, m.deghat@unsw.edu.au, j.guivant@unsw.edu.au).}

\begin{abstract}
This paper addresses bearing-only algorithms for solving the Fermat–Weber Location Problem (FWLP) with a unicycle agent. 
Unlike existing FWLP solutions for single- or double-integrator agents, our approach accounts for the nonholonomic constraints of wheeled robots. We first develop a bearing-only control law for the case with stationary beacons. Next, we consider saturated control inputs and propose a corresponding bearing-only control law. Finally, we address moving beacons with constant velocities and develop a control law that enables the unicycle agent to track the moving Fermat–Weber point. Both simulations and experiments are provided to demonstrate the effectiveness of the proposed methods.
\end{abstract}


\end{frontmatter}

\section{Introduction}
The Fermat-Weber Location Problem (FWLP) is an optimization problem that seeks a position that minimizes the weighted sum of Euclidean distances to a given set of beacons. This optimal position is also known as the Fermat–Weber point. The FWLP was first investigated by Endre Weiszfeld \cite[]{brimberg_1995_fermat,Plastria_2011_weiszfeld}. His original analysis considered only the unweighted case, assuming the weights of all beacons were equal to one. Extensions to the weighted case, using similar analytical techniques, can be found in \cite{Beck_2015_weiszfeld}. 

In practice, the FWLP is applicable to surveillance tasks, where the agent moves to a position that optimally supervises all beacons. In addition, it can be used for resource allocation. Although the Fermat-Weber point can be determined using active sensors such as LiDAR, small-sized agents typically rely on low-cost, lightweight passive onboard sensors to reduce hardware complexity and cost. In many guidance applications, bearing measurements are preferred because they are simple and can be easily obtained using light-weight cameras \cite[]{tron_2016_a}.

The term \textit{bearing-only} means that the control objective must be achieved using only bearing measurements, without relying on additional measurements such as distances, displacements, or relative velocities. For example, \cite{deghat_2014_localization} investigated bearing-only target localization and circumnavigation, in which an agent uses bearing measurements to estimate a target’s position and circumnavigate around it. \cite{duan_2023_bearing} studied bearing-only containment control for single-integrator agents. In containment control, agents are required to move into the convex hull formed by the beacons, but not the Fermat-Weber point. In \cite{zhao_2016_localizability,trinh_2021_robust,vantran_2022_bearing}, bearing-only formation tracking control has been addressed, allowing multiple agents to achieve a desired formation specified by bearings and the positions of the leaders. The primary distinction between bearing-only formation tracking and bearing-only FWLP is that the Fermat–Weber point is implicitly defined by an algebraic equation involving the beacons and cannot be directly computed from the positions of the beacons and bearing measurements. The agent must rely solely on bearing information to identify and track the point that satisfies this algebraic equation.

Bearing-only FWLP was first investigated in \cite{Trinh_2015_fermat}, where the authors considered stationary beacons and proposed a bearing-only control law to steer a single-integrator agent to the Fermat-Weber point. Additionally, \cite{LePhan_2025_bearing} extended this framework to accommodate double-integrator agents and moving beacons. More recently, \cite{Cheah_2025_bearing} studied the FWLP for Euler–Lagrange systems. Therefore, solving the bearing-only FWLP for unicycle agents remains unexplored.

In this work, we solve the bearing-only FWLP using a unicycle agent. Although the unicycle model can be approximated by a single-integrator model via feedback linearization \cite[]{Tnunay_2017_distributed}, this approximation may lead to control inputs with unrealistic magnitudes or trajectories that are infeasible due to the nonholonomic constraints of the unicycle model. Similarly, although the unicycle model can also be converted into a double-integrator model, this approach requires the velocity of the unicycle to be nonzero, which may not always hold in practice \cite[]{Liu_2013_distributed}. Hence, existing bearing-only solutions to the FWLP are not applicable to unicycle agents, and it is crucial to develop new bearing-only control laws specifically for the unicycle model to ensure practical feasibility. The contributions of this paper are summarized below.

\noindent($i$) We present bearing-only solutions to the FWLP for the unicycle agent. We begin by considering stationary beacons and propose a bearing-only control law that navigates the unicycle agent to the Fermat–Weber point. We further consider the case where the unicycle agent is subject to input constraints, which commonly arise in practice due to actuator saturation. To address this, we design a bearing-only control law that ensures the unicycle agent converges to the Fermat–Weber point despite these limitations. Experimental results are also provided to demonstrate that the proposed control law can be implemented on a real unicycle agent.

\noindent($ii$) We address the case where beacons move with a common constant velocity and propose a bearing-only control law that allows the unicycle agent to track the moving Fermat–Weber point.


\noindent\textit{Notations:} Let $\|\cdot\|$ be the Euclidean norm of a vector. For $d\in\mathbb{N}$, let $\bm{0}_d = [0,...,0]^\top\in\mathbb{R}^d$. Let $I_d$ be the $d\times d$ identity matrix. Given $A_i\in \mathbb{R}^{p\times q}$, for $i=1,...,n$, denote $\text{diag}(A_i):=\text{blkdiag}\{A_1,...,A_n\}\in\mathbb{R}^{pn\times qn}$. 

\section{Preliminaries}

\subsection{Unicycle Model}
Let $\bm{p}=[x,y]^\top\in\mathbb{R}^2$ and $\vartheta\in\mathbb{R}$ be the position and the heading angle of a unicycle agent, respectively. The unicycle model in 2D can be written as \cite[]{zhao_2018_affine}
\begin{equation} \label{e1}
    \dot{x} = \nu \cos(\vartheta),\quad 
    \dot{y} = \nu \sin(\vartheta),\quad 
    \dot{\vartheta} = \omega,
\end{equation}
where $\nu\in\mathbb{R}$ and $\omega\in\mathbb{R}$ are the linear speed and angular rate of the unicycle robot. Denote the vector of heading direction $\bm{h}$ as $\bm{h} = [\cos(\vartheta), \sin(\vartheta)]^\top$. The time derivative of $\bm{h}$ is
\begin{equation} \label{e2}
    \dot{\bm{h}} = \bm{h}^\perp \omega,
\end{equation}
where $\bm{h}^\perp = [-\sin(\vartheta),\cos(\vartheta)]^\top$. It can be verified that
\begin{equation}\label{e3}
    \bm{h}^\perp(\bm{h}^\perp)^\top = I_2-\bm{h}\bm{h}^\top.
\end{equation}
Then, the unicycle model can be rewritten as \cite[]{zhao_2018_affine} 
\begin{equation} \label{e4}
    \dot{\bm{p}} = \bm{h}{\nu},\quad
    \dot{\bm{h}} = \bm{h}^\perp \omega.
\end{equation}
Although this paper focuses on a unicycle agent moving in 2D, the model in \eqref{e4} can be extended to nonholonomic agents in 3D space, as explained in \cite{zhao_2018_affine}.

\subsection{Fermat-Weber Location Problem}
Let the positions of the beacons be $\bm{p}_i\in\mathbb{R}^2$, where $i=1,2,...,n$ and $n\geq3$ is the total number of beacons. The beacons are assumed to be non-collinear. Our aim is to design bearing-only control laws for $\nu$ and $\omega$ such that the unicycle agent reaches the Fermat–Weber point $\bm{p}^*\in\mathbb{R}^2$, which satisfies \cite[]{Plastria_2011_weiszfeld}
\begin{equation} \label{e5}
    \bm{p}^* = \underset{\bm{p}} {\mathrm{argmin}} \sum_{i=1}^n \gamma_i \|\bm{p}-\bm{p}_i\|,
\end{equation}
where $\gamma_i > 0$ are positive weights of the beacons.

Define the displacement vector $\bm{e}_i\in\mathbb{R}^2$ and the bearing vector $\bm{g}_i\in\mathbb{R}^2$ as
\begin{align}
    \bm{e}_i&:= \bm{p}_i-\bm{p}, \label{e6}\\ 
    \bm{g}_i&:= \frac{\bm{e}_i}{\|\bm{e}_i\|}. \label{e7}
\end{align}
Moreover, define the orthogonal projection matrix $P_{\bm{g}_i}$ as
\begin{equation} \label{e8}
    P_{\bm{g}_i}:=I_2-\bm{g}_i\bm{g}_i^\top.
\end{equation}
Note that $P_{\bm{g}_i}$ is a positive semidefinite matrix with $\text{Null}(P_{\bm{g}_i})=\text{span}\{\bm{g}_i\}$ \cite[]{zhao_2016_bearing}.
The following lemma is well-known in the FWLP literature.

\noindent\textbf{Lemma 1 \cite[]{Plastria_2011_weiszfeld}.} \textit{A unique Fermat–Weber point  $\bm{p}^*$ exists if the following inequality }
\begin{equation} \label{e9}
    \left\|\sum_{i=1;i\neq k}^n \gamma_i \frac{\bm{p}_i-\bm{p}_k}{\|\bm{p}_i-\bm{p}_k\|}\right\| >\gamma_k,\quad k=1,...,n,
\end{equation}
\textit{holds. Furthermore, if there exists a point $\mathbf{m}\in\mathbb{R}^2$ different from all $\bm{p}_i,\ i=1,...,n$, for which the following equation holds}
\begin{equation} \label{e10}
    \sum_{i=1}^n \gamma_i\frac{\bm{p}_i-\mathbf{m}}{\|\bm{p}_i-\mathbf{m}\|}=\bm{0}_2,
\end{equation} 
\textit{then $\mathbf{m}$ is the Fermat-Weber point $\bm{p}^*$.}

Let $\bm{g}^*_i=\frac{\bm{p}_i-\bm{p}^*}{\|\bm{p}_i-\bm{p}^*\|}$, for $i=1,2,...,n$. From \eqref{e10}, it can be verified that $\sum_{i=1}^n \gamma_i\bm{g}_i^*=\bm{0}_2$.




\section{Bearing-only solution to the FWLP for unicycle agent: stationary beacons}

\subsection{Problem Formulation}
The following assumptions are introduced and will be used in the stability analysis.

\noindent\textit{Assumption 1. The unicycle agent can measure the relative bearing $\bm{g}_i$ to the beacons. } \\
\noindent\textit{Assumption 2. No collisions occur between the unicycle agent and the beacons.}

Assumption 2 serves as a simplifying assumption for the bearing-only FWLP, as $\bm{g}_i$ becomes undefined when the agent collides with a beacon. Similar to \cite[Lemma 5]{LePhan_2025_bearing}, a sufficient condition for collision avoidance can be established. Under Assumptions 1-2, the following problem is introduced.

\noindent\textbf{Problem 1.} \textit{ Consider the unicycle agent in \eqref{e1} and assume the beacons are stationary. Design a bearing-only control law such that the unicycle agent reaches the Fermat-Weber point as $t\rightarrow\infty$.}

\subsection{Proposed Control Law}
The following bearing-only control law is proposed to address Problem 1:
\begin{equation} \label{e11}
    \nu = k_p\bm{h}^\top\sum_{i=1}^n\gamma_i\bm{g}_i, \quad
    \omega =k_h (\bm{h}^\perp)^\top\sum_{i=1}^n\gamma_i\bm{g}_i,
\end{equation}
where $k_p$ and $k_h$ are positive control gains. Substituting \eqref{e11} into \eqref{e4} yields
\begin{align*}
    \dot{\bm{p}} &= k_p \bm{h} \bm{h}^\top\sum_{i=1}^n\gamma_i\bm{g}_i,\numberthis \label{e12}\\
    \dot{\bm{h}} &= k_h\bm{h}^\perp (\bm{h}^\perp)^\top\sum_{i=1}^n\gamma_i\bm{g}_i\stackrel{\eqref{e3}}{=} k_h(I_2-\bm{h}\bm{h}^\top)\sum_{i=1}^n\gamma_i\bm{g}_i. \numberthis \label{e13}
\end{align*}

\subsection{Stability Analysis}
Denote $\bm{e}=[\bm{e}_1^\top,...,\bm{e}_n^\top]^\top$, $\bm{g}=[\bm{g}_1^\top,...,\bm{g}_n^\top]^\top$, $\bm{g}^*=[\bm{g}_1^{*\top},...,\bm{g}_n^{*\top}]^\top$ and $W=\text{diag}(\gamma_iI_2)$. The following theorem demonstrates that, under the proposed control law \eqref{e11}, the unicycle agent asymptotically converges to the Fermat–Weber point $\bm{p}^*$. 

\noindent\textbf{Theorem 1.} \textit{Under Assumptions 1-2 and assuming that a unique Fermat-Weber point $\bm{p}^*$ exists according to Lemma 1, the unicycle agent \eqref{e1} driven by the control law \eqref{e11} converges to $\bm{p}^*$ as $t\rightarrow\infty$.}

\noindent\textit{Proof.} Consider the following Lyapunov function candidate
\begin{equation} \label{e14}
    V_1 = \bm{e}^\top W(\bm{g}-\bm{g}^*).
\end{equation}
The function $V_1$ is strictly positive and equals zero if and only if $\bm{g}-\bm{g}^*=\bm{0}_{2n}$, which implies $\bm{p}=\bm{p}^*$ \cite[Lemma 2]{LePhan_2025_bearing}. Taking the time derivative of \eqref{e14} yields
\begin{align*}
    \dot{V}_1 &= (\bm{g}-\bm{g}^*)^\top W \dot{\bm{e}}+\bm{e}^\top W\dot{\bm{g}}.
\end{align*}
The second term of $\dot{V}_1$ is zero as $\bm{e}^\top W\dot{\bm{g}}=\sum_{i=1}^n\gamma_i \bm{e}_i^\top \dot{\bm{g}}_i\stackrel{\eqref{e7}}{=}\sum_{i=1}^n\gamma_i\bm{e}_i^\top (\dot{\bm{e}}_i\|\bm{e}_i\|-\bm{e}_i\frac{\mathrm{d}}{\mathrm{d}t}\|\bm{e}_i\|)/\|\bm{e}_i\|^2$. Since $\frac{\mathrm{d}}{\mathrm{d}t}\|\bm{e}_i\|=\bm{e}_i^\top\dot{\bm{e}}_i/\|\bm{e}_i\|=\bm{g}_i^\top\dot{\bm{e}}_i$, we have
\begin{align*}
    \bm{e}^\top W\dot{\bm{g}}&=\sum_{i=1}^n\gamma_i\bm{e}_i^\top\frac{(I_2-\bm{g}_i\bm{g}_i^\top)\dot{\bm{e}}_i}{\|\bm{e}_i\|}\stackrel{\eqref{e8}}{=}\sum_{i=1}^n\gamma_i\bm{g}_i^\top P_{\bm{g}_i}\dot{\bm{e}}_i =0.
\end{align*}
Therefore, $\dot{V}_1$ can be written as
\begin{align*}
    \dot{V}_1&\stackrel{\eqref{e6}}{=}-\sum_{i=1}^n \gamma_i(\bm{g}_i-\bm{g}_i^*)^\top \dot{\bm{p}}\\
    &\stackrel{\eqref{e12}}{=} -k_p \sum_{i=1}^n \gamma_i(\bm{g}_i-\bm{g}^*)^\top \bm{h}\bm{h}^\top \left(\sum_{i=1}^n\gamma_i\bm{g}_i\right)\\
    &\stackrel{\eqref{e10}}{=} -k_p \left\| \bm{h}^\top\sum_{i=1}^n \gamma_i \bm{g}_i \right\|^2\leq 0. \numberthis \label{e15}
\end{align*}
Applying LaSalle's invariance principle, it follows that $\bm{h}^\top\sum_{i=1}^n \gamma_i \bm{g}_i\rightarrow 0$ as $t\rightarrow\infty$. The term $\bm{h}^\top\sum_{i=1}^n \gamma_i \bm{g}_i=0$ implies that either ($i$) $\sum_{i=1}^n \gamma_i \bm{g}_i=\bm{0}_2$ or ($ii$) $\bm{h} \perp \sum_{i=1}^n \gamma_i \bm{g}_i$. The condition in case ($ii$) cannot hold because according to \eqref{e13}, when $\bm{h} \perp \sum_{i=1}^n \gamma_i \bm{g}_i$, the vector $\dot{\bm{h}}$ is nonzero, which implies that the vector $\bm{h}$ changes. As a result, $\bm{h}$ is no longer perpendicular to $\sum_{i=1}^n \gamma_i \bm{g}_i$. Therefore, case $(i)$ is the only possible solution to satisfy $\dot{V}_1=0$. From \eqref{e10} in Lemma 1, the unicycle reaches the Fermat-Weber point when $\sum_{i=1}^n \gamma_i \bm{g}_i=\bm{0}_2$. $\hfill \blacksquare$

\subsection{Bearing-Only Solution to the FWLP Subject to \\Constraints}

Now, suppose $\nu$ and $\omega$ in \eqref{e1} are constrained by
\begin{align}
    -\nu_b\leq \nu \leq \nu_f,\\
    -\omega_r \leq \omega\leq \omega_l,
\end{align}
where $\nu_b > 0$ and $\nu_f > 0$ denote the maximum backward and forward linear speeds, respectively, while $\omega_l > 0$ and $\omega_r > 0$ represent the maximum angular rates for left and right turns, respectively. Define the following saturation functions
for the linear speed and angular rate as
\begin{align}
    \text{sat}_\nu(x) =\begin{cases} -\nu_b, &x\in(-\infty,-\nu_b),\\
    x, &x\in[-\nu_b,\nu_f],\\
    \nu_f, &x\in(\nu_f,\infty),
    \end{cases} \\
    \text{sat}_\omega(x) =\begin{cases} -\omega_r, &x\in(-\infty,-\omega_r),\\
    x, &x\in[-\omega_r,\omega_l],\\
    \omega_l, &x\in(\omega_l,\infty).
    \end{cases} 
\end{align}
We propose the bearing-only control law for the unicycle agent as
\begin{equation}\label{e20}
    \nu =\text{sat}_\nu\left(\bm{h}^\top \sum_{i=1}^n\gamma_i\bm{g}_i\right),\quad
    \omega = \text{sat}_\omega \left((\bm{h}^\perp)^\top \sum_{i=1}^n\gamma_i\bm{g}_i\right).
\end{equation}
The following theorem studies the stability of the unicycle agent controlled by \eqref{e20}. 

\noindent\textbf{Theorem 2.} \textit{Under Assumptions 1-2 and assuming that there exists a unique Fermat-Weber point $\bm{p}^*$ according to Lemma 1, the unicycle agent \eqref{e1} driven by the control law \eqref{e20} converges to  $\bm{p}^*$ as $t\rightarrow\infty$.}

\noindent\textit{Proof.} The stability analysis follows a similar approach to that in \cite[Theorems 3-4]{zhao_2018_ageneral}. Substituting \eqref{e20} in \eqref{e4} yields
\begin{equation} \label{e21}
\begin{split}
    \dot{\bm{p}}&= \bm{h} \text{sat}_\nu\left(\bm{h}^\top \sum_{i=1}^n\gamma_i\bm{g}_i\right),\\
    \dot{\bm{h}}&= \bm{h}^\perp \text{sat}_\omega \left((\bm{h}^\perp)^\top \sum_{i=1}^n\gamma_i\bm{g}_i\right).
\end{split}
\end{equation}
The saturation function can be rewritten as\\ $\mathrm{sat}_\nu\left(\bm{h}^\top \sum_{i=1}^n\gamma_i\bm{g}_i\right)=\kappa\bm{h}^\top \sum_{i=1}^n\gamma_i\bm{g}_i$, where 
\begin{equation}
    \kappa = \begin{cases}
        -\frac{\nu_b}{\bm{h}^\top \sum_{i=1}^n\gamma_i\bm{g}_i}, & \bm{h}^\top \sum_{i=1}^n\gamma_i\bm{g}_i\in(-\infty,-\nu_b)\\
        1,& \bm{h}^\top \sum_{i=1}^n\gamma_i\bm{g}_i\in[-\nu_b,\nu_f]\\
        \frac{\nu_f}{\bm{h}^\top \sum_{i=1}^n\gamma_i\bm{g}_i}, &\bm{h}^\top \sum_{i=1}^n\gamma_i\bm{g}_i\in(\nu_f,\infty).
    \end{cases}   
\end{equation}
Observe that $\kappa$ is always positive and continuous, but not smooth. Moreover, rewrite $\text{sat}_\omega \left((\bm{h}^\perp)^\top \sum_{i=1}^n\gamma_i\bm{g}_i\right)=\rho(\bm{h}^\perp)^\top \sum_{i=1}^n\gamma_i\bm{g}_i$, where
\begin{equation}\label{e23}
        \rho = \begin{cases}
        -\frac{\omega_r}{(\bm{h}^\perp)^\top \sum_{i=1}^n\gamma_i\bm{g}_i}, & (\bm{h}^\perp)^\top \sum_{i=1}^n\gamma_i\bm{g}_i\in(-\infty,-\omega_r)\\
        1,&(\bm{h}^\perp)^\top \sum_{i=1}^n\gamma_i\bm{g}_i\in[-\omega_r,\omega_l]\\
        \frac{\omega_l}{(\bm{h}^\perp)^\top \sum_{i=1}^n\gamma_i\bm{g}_i}, &(\bm{h}^\perp)^\top \sum_{i=1}^n\gamma_i\bm{g}_i\in(\omega_l,\infty).
        \end{cases}
\end{equation}
As a result, $\dot{\bm{h}}=\bm{h}^\perp(\bm{h}^\perp)^\top \rho \sum_{i=1}^n\gamma_i\bm{g}_i=\\(I_2-\bm{h}\bm{h}^\top)(\rho\sum_{i=1}^n\gamma_i\bm{g}_i)$. Using the same Lyapunov function \eqref{e14}, the time derivative of $V_1$ is
\begin{align*}
    \dot{V}_1&=-\sum_{i=1}^n \gamma_i(\bm{g}_i-\bm{g}_i^*)^\top \dot{\bm{p}}\\
    &\stackrel{\eqref{e21}}{=}-\kappa \sum_{i=1}^n \gamma_i(\bm{g}_i-\bm{g}_i^*)^\top \bm{h} \bm{h}^\top \sum_{i=1}^n\gamma_i\bm{g}_i \\
    &\stackrel{\eqref{e10}}{=} -\kappa \left\| \bm{h}^\top\sum_{i=1}^n \gamma_i \bm{g}_i \right\|^2\leq 0. \numberthis\label{e24}
\end{align*}
Hence, $V_1$ is bounded and converges as $t\rightarrow\infty$. The next step is to show that $\dot{V}_1$ is uniformly continuous in $t$ by proving that $\kappa$, $\bm{h}$, and $\sum_{i=1}^n \gamma_i \bm{g}_i$ are uniformly continuous in $t$. Since the proposed control law \eqref{e20} is built upon the unified controller introduced in \cite{zhao_2018_ageneral}, the remainder of the proof follows directly from the arguments in \cite[Theorems 3–4]{zhao_2018_ageneral} and we can show that $\sum_{i=1}^n \gamma_i \bm{g}_i\rightarrow \bm{0}_2$ as $t\rightarrow\infty$. Using Lemma 1, since $\sum_{i=1}^n \gamma_i \bm{g}_i\rightarrow \bm{0}_2$ as $t\rightarrow\infty$, we have $\bm{p}-\bm{p}^*\to\bm{0}_2$.  $\hfill \blacksquare$

\section{Bearing-only solution to the FWLP for unicycle agent: moving beacons}
\subsection{Problem Formulation}
Let $\bm{v}^*\in\mathbb{R}^2$ be the constant velocity of the beacons. Given Assumptions 1 and 2, the following problem is formulated.

\noindent\textbf{Problem 2.} \textit{Consider the unicycle agent described by \eqref{e1} and a set of beacons moving with constant velocity $\bm{v}^*$. Design a bearing-only control law such that the unicycle agent goes to the Fermat–Weber point as $t\rightarrow\infty$.}

\subsection{Proposed Control Law}
The following bearing-only control law is proposed to address Problem 2:
{\small
\begin{equation} \label{e25}
\begin{split}
    \nu &= \bm{h}^\top\left(k_1 \sum_{i=1}^n\gamma_i\bm{g}_i + \bm{\phi} \right),\
    \omega = k_2 (\bm{h}^\perp)^\top \left(\sum_{i=1}^n\gamma_i\bm{g}_i+\bm{\phi}\right),  \\
    \dot{\bm{\phi}} &= k_3\left( \bm{h}\bm{h}^\top\left(\sum_{i=1}^n\gamma_i\bm{g}_i\right) - (I_2-\bm{h}\bm{h}^\top)\bm{\phi}\right),
\end{split}    
\end{equation}
}
where $k_1$, $k_2$ and $k_3$ are positive control gains. Substituting \eqref{e25} into \eqref{e4} yields
\begin{equation} \label{e26}
\begin{split}
    \dot{\bm{p}} &= \bm{h} \bm{h}^\top\left(k_1 \sum_{i=1}^n\gamma_i\bm{g}_i + \bm{\phi} \right),\\
    \dot{\bm{h}} &\stackrel{\eqref{e3}}{=} k_2 (I_2-\bm{h}\bm{h}^\top) \left(\sum_{i=1}^n\gamma_i\bm{g}_i+\bm{\phi}\right),  \\
    \dot{\bm{\phi}} &= k_3\left( \bm{h}\bm{h}^\top\left(\sum_{i=1}^n\gamma_i\bm{g}_i\right) - (I_2-\bm{h}\bm{h}^\top)\bm{\phi}\right).
\end{split}    
\end{equation}

\subsection{Stability Analysis}
The following theorem shows that the unicycle agent under the proposed control law \eqref{e25} asymptotically converges to the moving Fermat–Weber point $\bm{p}^*$.

\noindent\textbf{Theorem 3.} \textit{Under Assumptions 1-2 and assuming that the beacons move with constant velocity $\bm{v}^*$, the unicycle agent governed by the control law \eqref{e25} converges to the Fermat-Weber point described in Lemma 1, that is, $\|\bm{p} - \bm{p}^*\| \rightarrow 0$ as $t \to \infty$.}

\noindent\textit{Proof.} Consider the following Lyapunov function candidate
\begin{equation}
    V_2 = \bm{e}^\top W(\bm{g}-\bm{g}^*) +\frac{1}{2k_3} \|\bm{\phi}-\bm{v}^*\|^2+\frac{\|\bm{v}^*\|}{2k_2}\|\bm{h}-\bm{h}^*\|^2, 
\end{equation}
where $\bm{h}^*=\frac{\bm{v}^*}{\|\bm{v}^*\|}\in\mathbb{R}^2$ is the desired heading direction of the unicycle which is constant. The time derivative of $V_2$ is
\begin{align*}
    \dot{V}_2&=-\sum_{i=1}^n \gamma_i(\bm{g}_i-\bm{g}_i^*)^\top \dot{\bm{p}}+\sum_{i=1}^n \gamma_i(\bm{g}_i-\bm{g}_i^*)^\top \bm{v}^*\\
    &\quad+\frac{1}{k_3}(\bm{\phi}-\bm{v}^*)^\top\dot{\bm{\phi}} +\frac{\|\bm{v}^*\|}{k_2}(\bm{h}-\bm{h}^*)^\top\dot{\bm{h}}\\
    &\stackrel{\eqref{e10},\eqref{e26}}{=} -k_1 \left\| \bm{h}^\top\sum_{i=1}^n \gamma_i \bm{g}_i \right\|^2 - \sum_{i=1}^n \gamma_i \bm{g}_i^\top \bm{h}\bm{h}^\top\bm{\phi}\\
    &\quad + \sum_{i=1}^n \gamma_i \bm{g}_i^\top \bm{v}^*  +(\bm{\phi}-\bm{v}^*)^\top \bm{h}\bm{h}^\top\left(\sum_{i=1}^n\gamma_i\bm{g}_i\right) \\
    &\quad - (\bm{\phi}-\bm{v}^*)^\top(I_2-\bm{h}\bm{h}^\top)\bm{\phi}\\
    &\quad +{\|\bm{v}^*\|}(\bm{h}-\bm{h}^*)^\top (I_2-\bm{h}\bm{h}^\top) \left(\sum_{i=1}^n\gamma_i\bm{g}_i+\bm{\phi}\right) \\
    &= -k_1 \left\| \bm{h}^\top\sum_{i=1}^n \gamma_i \bm{g}_i \right\|^2 - \sum_{i=1}^n \gamma_i \bm{g}_i^\top \bm{h}\bm{h}^\top\bm{\phi}\\
    &\quad + \sum_{i=1}^n \gamma_i \bm{g}_i^\top \bm{v}^*  +(\bm{\phi}-\bm{v}^*)^\top \bm{h}\bm{h}^\top\left(\sum_{i=1}^n\gamma_i\bm{g}_i\right) \\ 
    &\quad - \bm{\phi}^\top(I_2-\bm{h}\bm{h}^\top)\bm{\phi}+\bm{v}^{*\top}(I_2-\bm{h}\bm{h}^\top)\bm{\phi}\\
    &\quad -\bm{v}^{*\top}(I_2-\bm{h}\bm{h}^\top) \left(\sum_{i=1}^n\gamma_i\bm{g}_i+\bm{\phi}\right) \numberthis\label{e28}
\end{align*}
where the last equality is due to $\|\bm{v}^*\|\bm{h}^{*\top}= \bm{v}^{*\top}$ and $\bm{h}^\top(I_2-\bm{h}\bm{h}^\top)=\bm0_{2}^\top$. By canceling the common terms in \eqref{e28}, we obtain
\begin{align*}
\dot{V}_2 \stackrel{\eqref{e3}}{=} -k_1 \left\| \bm{h}^\top\sum_{i=1}^n \gamma_i \bm{g}_i \right\|^2 -\|(\bm{h}^\perp)^\top \bm{\phi}\|^2. \numberthis \label{e29} 
\end{align*}
Hence, $\dot{V}_2$ is negative semidefinite. By LaSalle's invariance principle, the system trajectories converge to the largest invariant set, where $\dot{V}_2=0$. This implies that 
\begin{equation} \label{e30}
    \bm{h}^\top\sum_{i=1}^n \gamma_i \bm{g}_i = 0,
\end{equation}
and 
\begin{equation} \label{e31}
    (\bm{h}^\perp)^\top \bm{\phi} = 0. 
\end{equation}
Substituting \eqref{e30} and \eqref{e31} into \eqref{e26}, it follows that
\begin{align*}
    \dot{\bm{p}} &\stackrel{\eqref{e30}}{=} \bm{h}\bm{h}^\top \bm{\phi}= I_2\bm{\phi}-(I_2 - \bm{h}\bm{h}^\top)\bm{\phi}\\
    & \stackrel{\eqref{e3},\eqref{e31}}{=}\bm{\phi},\numberthis \label{e32}\\
    \dot{\bm{h}}&\stackrel{\eqref{e30},\eqref{e31}}{=} k_2\sum_{i=1}^n \gamma_i \bm{g}_i, \numberthis \label{e33}\\
    \dot{\bm{\phi}}&\stackrel{\eqref{e30},\eqref{e31}}{=} \bm{0}_2. \numberthis \label{e34}
\end{align*}
Thus, in steady state, the unicycle moves with a constant velocity $\bm{\phi}$, while its heading direction $\bm{h}$ also remains constant, as shown in \eqref{e32} and \eqref{e34}. This is because, once the velocity reaches a constant value, no further rotational motion occurs, and consequently, the unicycle maintains a fixed heading. Moreover, since the heading direction $\bm{h}$ is constant in steady state, it has a finite limit. As $\dot{\bm{h}}$, $\dot{\bm{g}}_i$ and $\dot{\bm{\phi}}$ are bounded, $\ddot{\bm{h}}$ is also bounded. By Barbalat's lemma \cite[]{sastry_2013_nonlinear}, we have $\dot{\bm{h}}\rightarrow \bm{0}_2$ as $t\rightarrow \infty$. This implies that $\sum_{i=1}^n \gamma_i\bm{g}_i\rightarrow \bm{0}_2$ in \eqref{e33} as $t\rightarrow \infty$ and the unicycle agent reaches the Fermat-Weber point. $\hfill \blacksquare$


\section{Simulation and Experimental Results}

\subsection{Simulation Results}
\begin{figure}[htb!]
     \begin{subfigure}[h]{0.49\textwidth}
     \centering
     \includegraphics[scale =0.4]{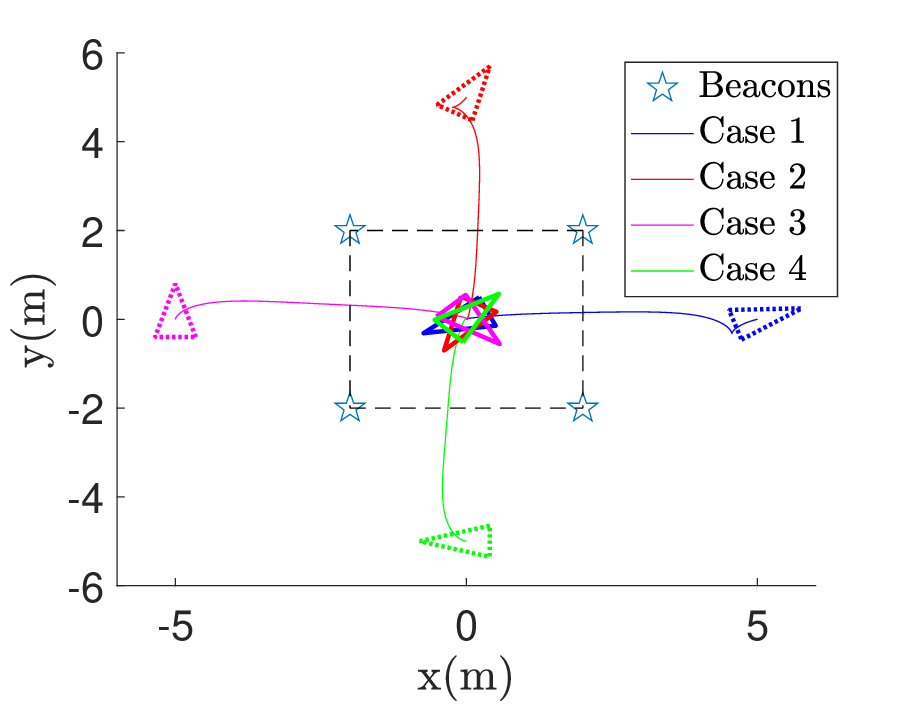}
     \caption{Unicycle agents trajectories.}
     \label{f1a}
     \end{subfigure}
     \begin{subfigure}[h]{0.49\textwidth}
     \centering
     \includegraphics[scale =0.4]{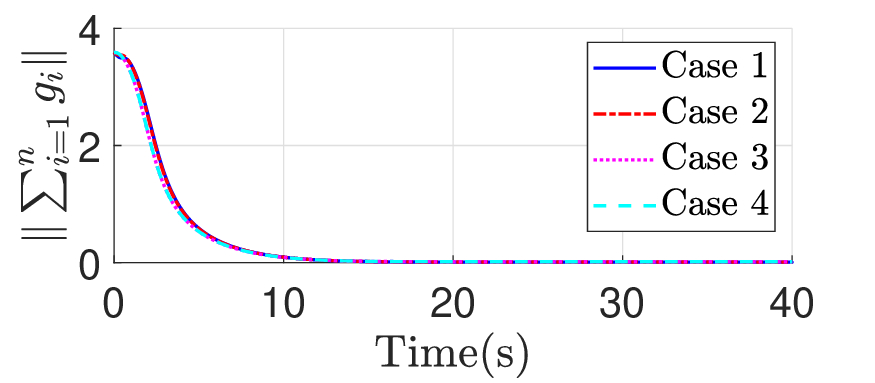}
     \caption{Tracking errors.}
     \label{f1b}
     \end{subfigure}
    \caption{Bearing-only FWLP of unicycle agents using the control law \eqref{e11}.}
    \label{f1}
\end{figure}

Fig. \ref{f1} shows the simulation of unicycle agents driven by the control law \eqref{e11}. The control parameters are set to $k_p=0.5$ and $k_h = 1$. There are four stationary beacons located at $\bm{p}_1 = [-2,2]^\top$, $\bm{p}_2 = [2,2]^\top$, $\bm{p}_3 = [2,-2]^\top$, and $\bm{p}_4 = [-2,-2]^\top$. The positive weights of the beacons are set to $\gamma_i=1$, for $i=1,...4$. 
It can also be verified, using \eqref{e9} in Lemma 1, that a unique Fermat–Weber point exists for this set of beacons. We consider four cases where a unicycle agent is placed at four different initial positions. As shown in Fig. \ref{f1}(a), they all converge to the Fermat–Weber point.

\begin{figure}[h]
     \begin{subfigure}[h]{0.49\textwidth}
     \centering
     \includegraphics[scale =0.4]{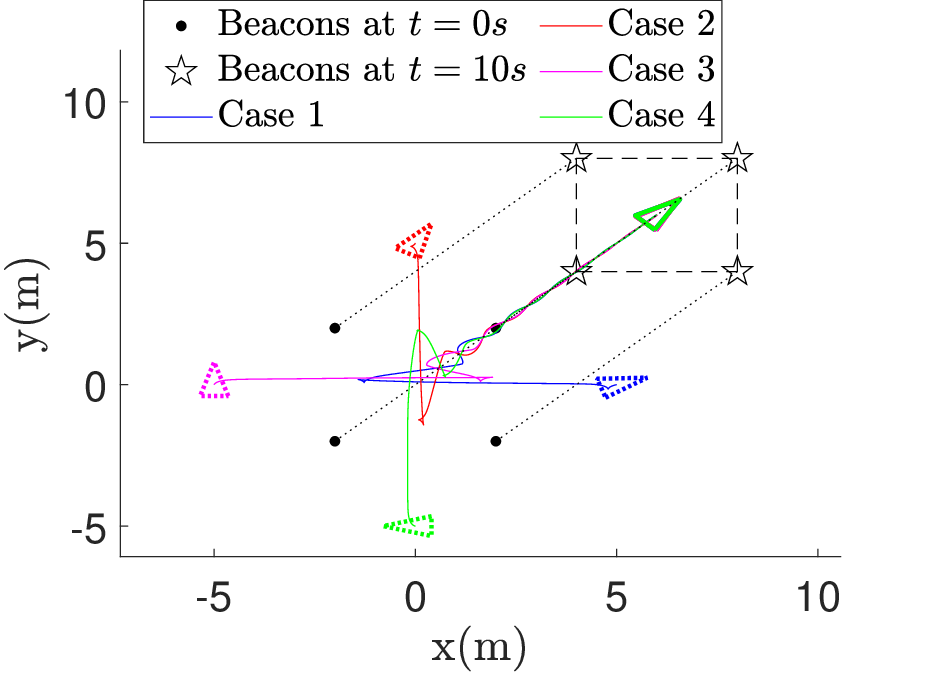}
     \caption{Unicycle agents trajectories.}
     \label{f2a}
     \end{subfigure}
     \begin{subfigure}[h]{0.49\textwidth}
     \centering
     \includegraphics[scale =0.4]{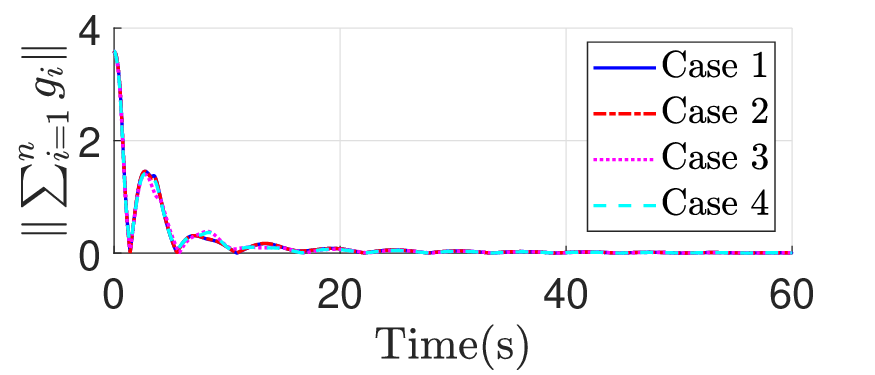}
     \caption{Tracking errors.}
     \label{f2b}
     \end{subfigure}
    \caption{Bearing-only FWLP of unicycle agents using the control law \eqref{e25}.}
    \label{f2}
\end{figure}

Fig. \ref{f2} presents the simulation results of a unicycle agent governed by the control law \eqref{e25}. The control parameters are set to $k_1 = 1$, $k_2 = 5$, and $k_3 = 1$. The beacons are initially positioned at $\bm{p}_1 = [-2,2]^\top$, $\bm{p}_2 = [2,2]^\top$, $\bm{p}_3 = [2,-2]^\top$, and $\bm{p}_4 = [-2,-2]^\top$, and move with a constant velocity of $\bm{v}^* = [0.1,0.1]^\top$. As shown in Fig. \ref{f2}(a), the unicycle agent initialized at different positions successfully track the moving Fermat–Weber point. 

\subsection{Experimental Results}
\begin{figure}[h]
     \begin{subfigure}[h]{0.49\textwidth}
     \centering
     \includegraphics[scale =0.1]{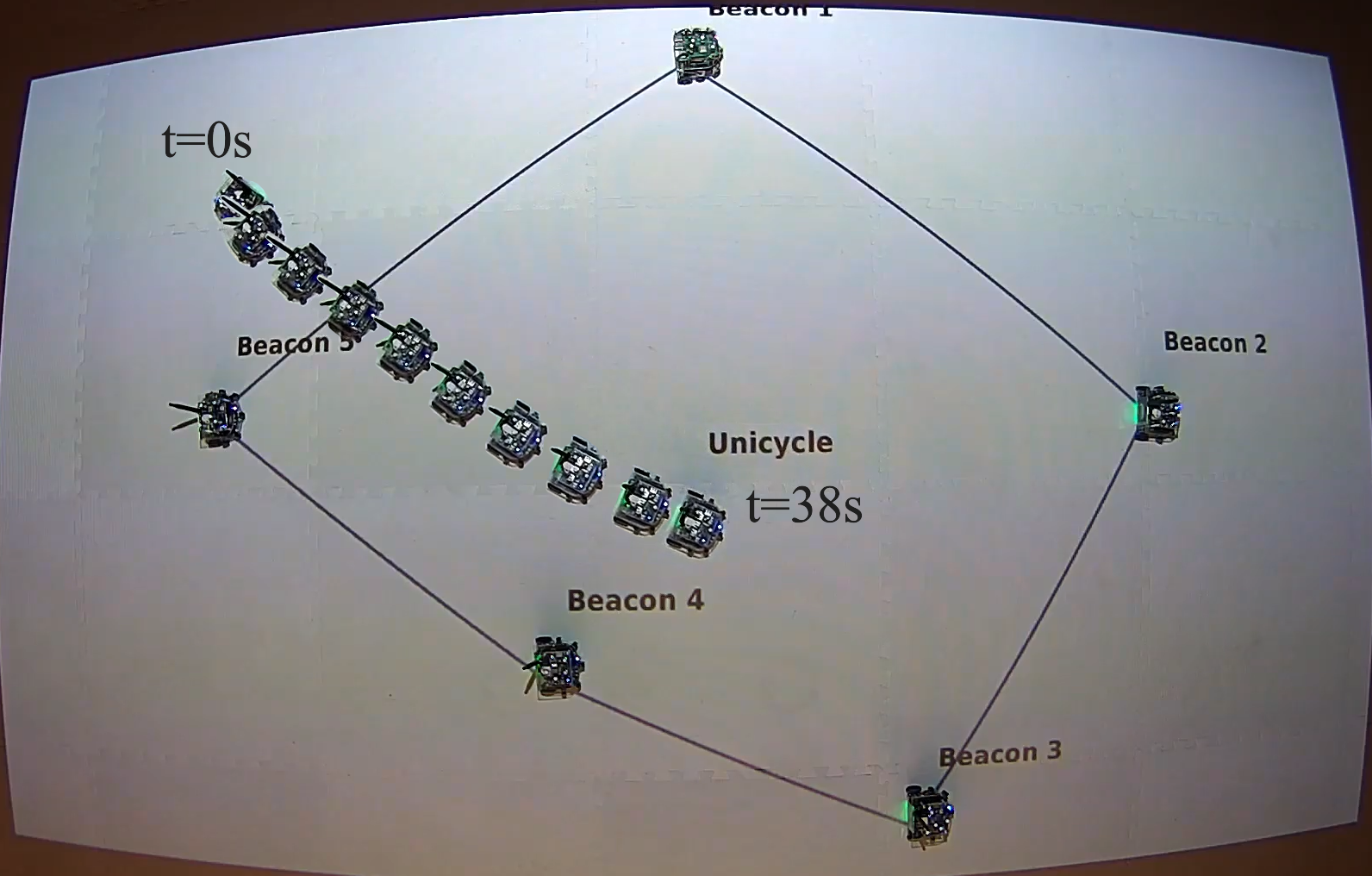}
     \caption{Unicycle agent trajectory.}
     \label{f3a}
     \end{subfigure}
     \begin{subfigure}[h]{0.49\textwidth}
     \centering
     \includegraphics[scale =0.4]{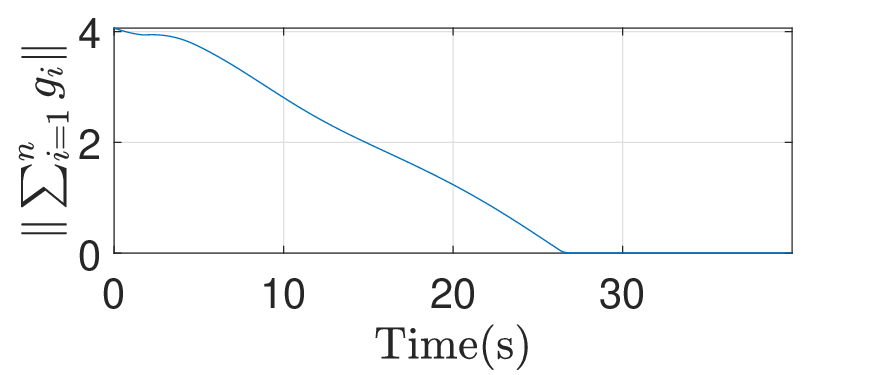}
     \caption{Tracking error.}
     \label{f3b}
     \end{subfigure}
    \begin{subfigure}[h]{0.49\textwidth}
     \centering
     \includegraphics[scale =0.4]{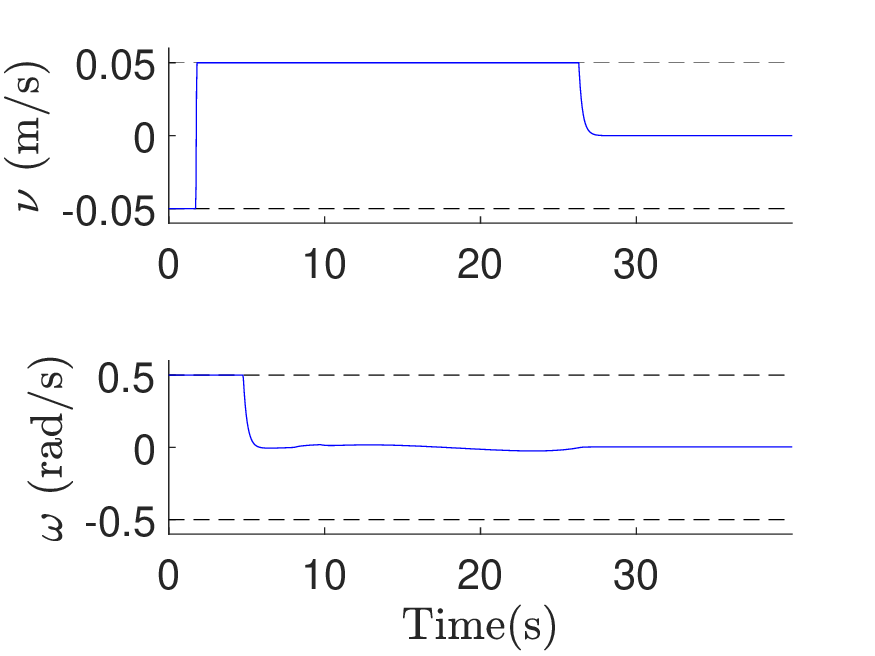}
     \caption{Linear speed and Angular speed.}
     \label{f3c}
     \end{subfigure}
    \caption{Experimental results of bearing-only FWLP using the control law \eqref{e20}.}
    \label{f3}
    \end{figure}

In this subsection, we validate the effectiveness of the proposed control laws \eqref{e20} and \eqref{e25} using the mobile robot provided by the Robotarium platform\footnote{The mobile robot is modeled by unicycle dynamics; see \cite[Eqn.~(S2)]{wilson_2020_robotarium}.} \cite[]{wilson_2020_robotarium}. Fig. \ref{f3} presents the experimental results of a unicycle robot controlled by \eqref{e20}. Linear speed and angular rate limits are set as $\nu_b = \nu_f = 0.05\text{m/s}$ and $\omega_r = \omega_l = 0.5\text{rad/s}$, respectively. We again assume that the beacons have equal weights. As shown in Fig. \ref{f3}, the unicycle robot successfully reached the Fermat–Weber point, under the imposed input constraints.

\begin{figure}[h]
     \begin{subfigure}[h]{0.49\textwidth}
     \centering
     \includegraphics[scale =0.23]{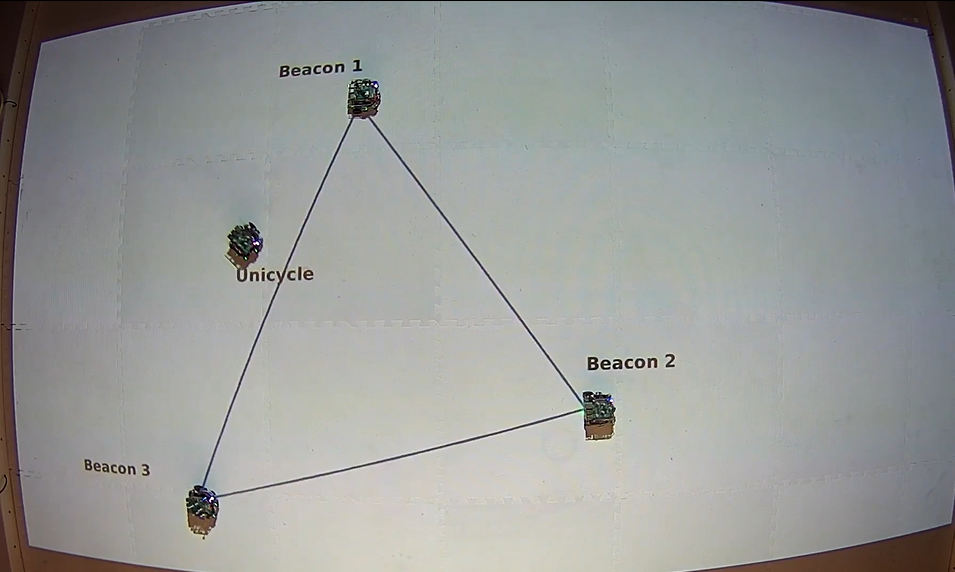}
     \caption{The unicycle agent and beacons positions at $t=0s$.}
     \label{f4a}
     \end{subfigure}
     \begin{subfigure}[h]{0.49\textwidth}
     \centering
     \includegraphics[scale =0.23]{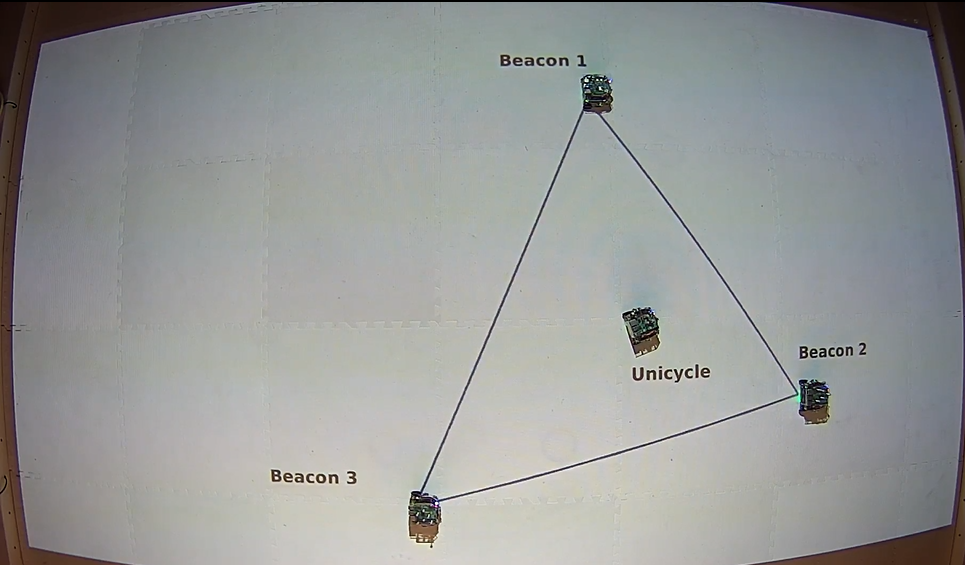}
     \caption{The unicycle agent and beacons positions at $t=39s$.}
     \label{f4b}
     \end{subfigure}
     \begin{subfigure}[h]{0.49\textwidth}
     \centering
     \includegraphics[scale =0.4]{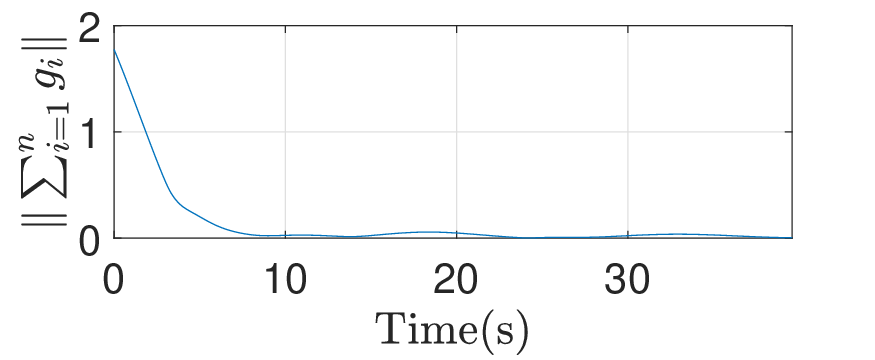}
     \caption{Tracking error.}
     \label{f4c}
     \end{subfigure}
    \caption{Experimental results of bearing-only FWLP using the control law \eqref{e25}.}
    \label{f4}
\end{figure}

Fig. \ref{f4} presents the experimental results of a unicycle robot controlled by \eqref{e25}. The control parameters are set to $k_1 = 0.5$, $k_2 = 4$, and $k_3 = 0.01$. In this experiment, the beacons are assumed to move with a constant velocity $\bm{v}^*=[0.05,0]^\top$. As shown in Fig. \ref{f4}, the unicycle agent successfully reached and tracked the moving Fermat–Weber point. The experimental video can be viewed at: \url{https://youtu.be/sksgUKmpl2Y}.

\section{Conclusion}
This paper studies the bearing-only solution to the FWLP for the unicycle agent. We first address the scenario in which the beacons are stationary and propose a bearing-only control law that ensures convergence of the unicycle to the Fermat-Weber point. We then extend the analysis to account for control input constraints. Furthermore, we propose a bearing-only control law for the case where the beacons move with a common constant velocity. Simulation and experimental results are provided to validate the proposed control laws. Future research may explore bearing-only solutions to the FWLP for the unicycle agent when beacons have negative weights, as well as the effect of measurement noise.

\bibliography{references}                                
\end{document}